\documentclass[preprint,amsmath,amssymb,11pt,aps,showpacs,showkeys]{revtex4}
\usepackage{graphicx}
\usepackage{graphics}
\usepackage{amsmath}
\usepackage{dcolumn}
\usepackage{amssymb}
\usepackage{bm}

\begin{document}
\title{Rotating effects on the scalar field in the cosmic string spacetime, in the spacetime with space-like dislocation and in the spacetime with a spiral dislocation}
\author{R. L. L. Vit\'oria}
\affiliation{Departamento de F\'isica, Universidade Federal da Para\'iba, Caixa Postal 5008, 58051-900, Jo\~ao Pessoa, PB, Brazil.}

\author{K. Bakke}
\email{kbakke@fisica.ufpb.br}
\affiliation{Departamento de F\'isica, Universidade Federal da Para\'iba, Caixa Postal 5008, 58051-900, Jo\~ao Pessoa, PB, Brazil.}

\begin{abstract}

In the interface between general relativity and relativistic quantum mechanics, we analyse rotating effects on the scalar field subject to a hard-wall confining potential. We consider three different scenarios of general relativity given by the cosmic string spacetime, the spacetime with space-like dislocation and the spacetime with a spiral dislocation. Then, by searching for a discrete spectrum of energy, we analyse analogues effects of the Aharonov-Bohm effect for bound states and the Sagnac effect.

\end{abstract}


\maketitle

\section{Introduction}

Two interesting points raised by Landau and Lifshitz \cite{landau3} are the effects of rotation in the Minkowski spacetime. One of them is the singular behaviour at larges distances for a system in a uniformly rotating frame, while the other is the effect on the clocks on the rotating body. In particular, this singular behaviour at larges distances means that there exists a restriction on the spatial coordinates due to the effects of rotation. For instance, by taking the line element of the Minkowski spacetime in cylindrical coordinates, we have: $ds^{2}=-c^{2}\,dt^{2}+dr^{2}+r^{2}\,d\varphi^{2}+dz^{2}$; thus, by taking $\varphi\rightarrow\varphi+\omega t$, where $\omega$ is the constant angular velocity of the rotating reference frame, then, the line element of the Minkowski spacetime becomes: $ds^{2}=-c^{2}\left(1-\omega^{2}\,r^{2}/c^{2}\right)dt^{2}+2\omega\,r^{2}d\varphi\,dt+dr^{2}+r^{2}\,d\varphi^{2}+dz^{2}$. Therefore, we can see that the radial coordinate becomes determined in the range: $0\,\leq\,r\,<c/\omega$. This restriction on the radial coordinate in a uniformly rotating frame has drawn attention to the interface between general relativity and relativistic quantum mechanics \cite{r4,r21,r24,b2}.

In recent years, aspects of the uniformly rotating frame have been investigated in the cosmic string spacetime \cite{b5}. It has been shown that the topology of the cosmic string spacetime also determines the restriction of the values of the radial coordinate. Then, rotating effects on relativistic quantum systems have been investigated in the background of the cosmic string spacetime, such as the Dirac oscillator \cite{b5}, the Klein-Gordon oscillator \cite{r20}, scalar bosons \cite{r22}, Duffin-Kemmer-Petiau equation \cite{r23} and nonrelativistic topological quantum scattering \cite{mb2}. In view of these studies, a point that has not been raised is the aspect of the scalar field inside a cylindrical shell, where rotation is present and the corresponding background is the cosmic string spacetime. In this interface between general relativity and quantum mechanics, it is interesting to search for the effects associated with rotation and the topology of the spacetime. Cosmic strings \cite{AV,vil,JSD,kibble} are linear topological defects in the spacetime characterized by a conical singularity \cite{staro}. This singularity is determined by the curvature concentrated on the symmetry axis of the cosmic string. Recently, a great deal of work has investigated the influence of the cosmic string spacetime on quantum systems \cite{corda,corda2,corda3,corda4,corda5,corda6,corda7,corda8,corda9,corda10,corda11,corda12,corda13} and in G\"odel-type spacetimes \cite{fur,fur2,valdir,vfb,godel}.

Another branch of research is topological defects in spacetime associated with torsion. A well-known example is the spacetime with a space-like dislocation \cite{put,valdir3}, which is an analogue of the screw dislocation in solids \cite{kat,kleinert,val}. This kind of topological defect background has been used in studies of the Aharonov-Bohm effect for bound states \cite{valdir2}, noninertial effects \cite{b}, relativistic position-dependent mass systems \cite{vb,vb2} and Kaluza-Klein theories \cite{fur4}. One point that has not been dealt with in the literature is the rotating effects on the scalar field by considering the cosmic string spacetime, the spacetime with space-like dislocation and the spacetime with a spiral dislocation as backgrounds. Therefore, in this work, we deal with a scalar field subject to a hard-wall confining potential in the cosmic string spacetime. We also consider a uniformly rotating frame. Then, we analyse a particular case where a discrete spectrum of energy can be obtained. We show that there exists the influence of the topology of the spacetime and rotation on the relativistic energy levels. Further, we consider two spacetime backgrounds with the presence of torsion \cite{put} and analyse rotating effects on the scalar field subject to a hard-wall confining potential.

The structure of this paper is: in section II, we start by introducing the line element of the cosmic string spacetime in a uniformly rotating frame. Then, we analyse the scalar field subject to a hard-wall confining potential; in section III, we obtain the line element of the spacetime with a space-like dislocation in a uniformly rotating frame, and thus, investigate the scalar field subject to a hard-wall confining potential; in section IV, we consider a spacetime with a spiral dislocation. Thus, we discuss the scalar field subject to a hard-wall confining potential in both nonrotating and uniformly rotating frame; in section V, we present our conclusions.

\section{Rotating effects in the cosmic string spacetime}

In studies of linear topological defects in the spacetime, the cosmic string is the most known example \cite{vil,kibble}. It is characterized by a line element that possesses a parameters related to the deficit of angle: $\alpha=1-4G\mu$, with $\mu$ as being the dimensionless linear mass density of the cosmic string and $G$ is the gravitational Newton constant. By working with the units $\hbar=c=1$ from now on, the line element of the cosmic string spacetime is written in the form:
\begin{eqnarray}
ds^{2}=-dt^{2}+dr^{2}+\alpha^{2}\,r^{2}\,d\varphi^{2}+dz^{2}.
\label{2.1}
\end{eqnarray}
Note that this topological defect has a curvature concentrated only on the cosmic string axis, i.e., the curvature tensor is given by $R_{\rho,\varphi}^{\rho,\varphi}=\frac{1-\alpha}{4\alpha}\,\delta_{2}(\vec{r})$, where $\delta_{2}(\vec{r})$ is the two-dimensional delta function \cite{staro}. Besides, the parameter $\alpha$ is defined in the range $0\,<\,\alpha\,<\,1$. In order to study the aspects of the uniformly rotating frame, let us perform a coordinate transformation given by $\varphi\rightarrow\varphi+\omega\,t$, where $\omega$ is is the constant angular velocity of the rotating frame. Thereby, the line element (\ref{2.1}) becomes \cite{b5}
\begin{eqnarray}
ds^{2}=-\left(1-\omega^{2}\alpha^{2}\,r^{2}\right)\,dt^{2}+2\omega\alpha^{2}\,r^{2}d\varphi\,dt+dr^{2}+\alpha^{2}r^{2}d\varphi^{2}+dz^{2}.
\label{2.3}
\end{eqnarray}

Hence, in the rotating reference frame, the line element (\ref{2.3}) show us that the radial coordinate in the cosmic string spacetime is restrict to the range:  
\begin{eqnarray}
0\,\leq\,r\,<\,\frac{1}{\omega\alpha},
\label{2.3a}
\end{eqnarray}
otherwise, we would have a particle placed outside of the light-cone. Note that by taking $\alpha\rightarrow1$, we recover the discussion made in Ref. \cite{landau3} in the Minkowski spacetime. Note that the restriction on the radial coordinate is determined by the angular velocity and by the parameter related to the deficit of angle.

Henceforth, let us study the effects of rotation on the scalar field in the cosmic string spacetime. In curved spacetime, the Klein-Gordon equation is written in the form \cite{mello2,vb}:
\begin{eqnarray}
m^{2}\phi=\frac{1}{\sqrt{-g}}\,\partial_{\mu}\left[\sqrt{-g}\,g^{\mu\nu}\,\partial_{\nu}\right]\phi,
\label{2.2}
\end{eqnarray}
where $g=\mathrm{det}\left(g_{\mu\nu}\right)$. Then, with the line element (\ref{2.3}), the Klein-Gordon equation (\ref{2.2}) becomes
\begin{eqnarray}
m^{2}\phi=-\frac{\partial^{2}\phi}{\partial t^{2}}+2\omega\,\frac{\partial^{2}\phi}{\partial\varphi\partial t}-\omega^{2}\,\frac{\partial^{2}\phi}{\partial\varphi^{2}}+\frac{\partial^{2}\phi}{\partial r^{2}}+\frac{1}{r}\,\frac{\partial\phi}{\partial r}+\frac{1}{\alpha^{2}r^{2}}\,\frac{\partial^{2}\phi}{\partial\varphi^{2}}+\frac{\partial^{2}\phi}{\partial z^{2}}.
\label{2.4}
\end{eqnarray}

With the cylindrical symmetry, we have that $\phi$ is an eigenfunction of the operators $\hat{p}_{z}=-i\partial_{z}$ and $\hat{L}_{z}=-i\partial_{\varphi}$. Therefore, a solution to Eq. (\ref{2.4}) can be given  in terms of the eigenvalues of the operator $\hat{p}_{z}$ and $\hat{L}_{z}$ as follows:
\begin{eqnarray}
\phi\left(t,\,\rho,\,\varphi,\,z\right)=e^{-i\,\mathcal{E}\,t}\,e^{i\,l\,\varphi}\,e^{i\,k\,z}\,f\left(r\right),
\label{2.5}
\end{eqnarray}
where $l=0,\pm1,\pm2,\ldots$ and $-\infty\,<\,k\,<\,\infty$. And then, by substituting the solution (\ref{2.5}) into Eq. (\ref{2.4}), we obtain the following radial equation:
\begin{eqnarray}
f''+\frac{1}{r}\,f'-\frac{l^{2}}{\alpha^{2}r^{2}}\,f+\lambda^{2}\,f=0,
\label{2.6}
\end{eqnarray}
where we have defined the parameter $\lambda$ in the form:
\begin{eqnarray}
\lambda^{2}=\left(\mathcal{E}+l\omega\right)^{2}-m^{2}-k^{2}.
\label{2.7}
\end{eqnarray}

Hence, Eq. (\ref{2.6}) is the well-known the Bessel differential equation \cite{abra}. The general solution to Eq. (\ref{2.6}) is given in the form: $f\left(r\right)=A\,J_{\frac{\left|l\right|}{\alpha}}\left(\lambda\,r\right)+B\,N_{\frac{\left|l\right|}{\alpha}}\left(\lambda\,r\right)$, where $J_{\frac{\left|l\right|}{\alpha}}\left(\lambda\,r\right)$ and $N_{\frac{\left|l\right|}{\alpha}}\left(\lambda\,r\right)$ are the Bessel function of first kind and second kind \cite{abra}, respectively. With the purpose of having a regular solution at the origin, we must take $B=0$ in the general solution since the Neumann function diverges at the origin. Thus, the regular solution to Eq. (\ref{2.6}) at the origin is given by: 
\begin{eqnarray}
f\left(r\right)=A\,J_{\frac{\left|l\right|}{\alpha}}\left(\lambda\,r\right). 
\label{2.8}
\end{eqnarray}

As we have pointed out in Eq. (\ref{2.3a}), the restriction on the radial coordinate imposes that the scalar field must vanish at $r\rightarrow\,\rho_{0}=1/\alpha\omega$. This means that the radial wave function (\ref{2.8}) must satisfy the boundary condition:
\begin{eqnarray}
f\left(r\rightarrow\,\rho_{0}=1/\alpha\omega\right)=0.
\label{2.9}
\end{eqnarray}
This corresponds to the scalar field subject to a hard-wall confining potential. In other words, the geometry of the spacetime plays the role of a hard-wall confining potential \cite{b5}. Next, let us consider a particular case where $\lambda\,\rho_{0}\gg1$. In this particular case, we can write \cite{abra,b}:
\begin{eqnarray}
J_{\frac{\left|l\right|}{\alpha}}\left(\lambda\,\rho_{0}\right)\rightarrow\sqrt{\frac{2}{\pi\lambda\,\rho_{0}}}\,\cos\left(\lambda\,\rho_{0}-\frac{\left|l\right|\pi}{2\alpha}-\frac{\pi}{4}\right).
\label{2.10}
\end{eqnarray}

Hence, by substituting (\ref{2.10}) into (\ref{2.8}), we obtain from the boundary condition (\ref{2.9}) that
\begin{eqnarray}
\mathcal{E}_{n,\,l,\,k}\approx -l\omega\pm \sqrt{m^{2}+\alpha^{2}\omega^{2}\pi^{2}\left[n+\frac{\left|l\right|}{2\alpha}\,+\frac{3}{4}\right]^{2}+k^{2}}.
\label{2.11}
\end{eqnarray}

Equation (\ref{2.11}) gives us the spectrum of energy of a scalar field subject to a hard-wall confining potential determined by the topology of the cosmic string spacetime in a uniformly rotating frame. The contributions to the relativistic energy levels (\ref{2.11}) that stem from the topology of the cosmic string are given by the effective angular momentum given by $l_{\mathrm{eff}}=\frac{\left|l\right|}{\alpha}$, and by the presence of the fixed radius $\rho_{0}=\frac{1}{\alpha\omega}$. Since there is no interaction between the scalar field and the topological defect, the presence of the effective angular momentum in the relativistic energy levels means that there exists an analogue of the Aharonov-Bohm effect for bound states \cite{pesk,valdir2,fur4,ab}. Besides, we can observe a Sagnac-type effect \cite{sag,r4,sag2,sag5} by the presence of the coupling between the angular velocity $\omega$ and angular momentum quantum number $l$. Furthermore, by taking the limit $\alpha\rightarrow1$, we have that $\rho_{0}\rightarrow\frac{1}{\omega}$ and the allowed energies of the system (\ref{2.11}) become
\begin{eqnarray}
\mathcal{E}_{n,\,l,\,k}\approx -l\omega\pm \sqrt{m^{2}+\omega^{2}\pi^{2}\left[n+\frac{\left|l\right|}{2}\,+\frac{3}{4}\right]^{2}+k^{2}},
\label{2.12}
\end{eqnarray}
which corresponds to the allowed energies of the scalar field subject to a hard-wall confining potential in the Minkowski spacetime in a uniformly rotating frame.

\section{Rotating effects in the spacetime with a space-like dislocation}

In Refs. \cite{put}, examples of topological defects in the spacetime associated with torsion are given. We start this section by considering a spacetime with a space-like dislocation, whose line element is given by 
\begin{eqnarray}
ds^{2}=-dt^{2}+dr^{2}+r^{2}\,d\varphi^{2}+\left(dz+\chi\,d\varphi\right)^{2},
\label{3.1}
\end{eqnarray}
where $\chi$ is a constant ($\chi>0$) associated with a dislocation in the spacetime \footnote{Note that the spatial part of the line element (\ref{3.1}) is known in the context of Condensed Matter Physics as screw dislocation \cite{val,valdir2,b}. }. Next, let us follow the previous section and make the coordinate transformation $\varphi\rightarrow\varphi+\omega\,t$. Then, the line element (\ref{3.1}) becomes \cite{b}:
\begin{eqnarray}
ds^{2}&=&-\left(1-\omega^{2}\,r^{2}-\chi^{2}\omega^{2}\right)\,dt^{2}+\left(2\omega\,r^{2}+2\chi^{2}\omega\right)d\varphi\,dt+dr^{2}+r^{2}d\varphi^{2}\nonumber\\
&+&2\chi\omega\,dz\,dt+\left(dz+\chi\,d\varphi\right)^{2}.
\label{3.2}
\end{eqnarray}
Hence, we can see that the radial coordinate is restricted to the range:
\begin{eqnarray}
0\,\leq\,r\,<\frac{\sqrt{1-\chi^{2}\omega^{2}}}{\omega}.
\label{3.3}
\end{eqnarray}
Observe that the restriction on the radial coordinate is determined by the angular velocity and the parameter associated with the torsion of the defect in contrast to that given in Eq. (\ref{2.3a}) for the cosmic string spacetime. In this case, if $r\,\geq\,\frac{\sqrt{1-\chi^{2}\omega^{2}}}{\omega}$ we would have a particle is placed outside of the light-cone. Note that by taking $\chi=0$, we recover the discussion made in Ref. \cite{landau3} in the Minkowski spacetime.

Let us go further by writing the Klein-Gordon equation (\ref{2.2}) in the spacetime described by the line element (\ref{3.2}):
\begin{eqnarray}
m^{2}\,\phi=-\frac{\partial^{2}\phi}{\partial t^{2}}+2\omega\,\frac{\partial^{2}\phi}{\partial\varphi\,\partial t}-\omega^{2}\,\frac{\partial^{2}\phi}{\partial\varphi^{2}}+\frac{\partial^{2}\phi}{\partial\rho^{2}}+\frac{1}{\rho}\frac{\partial\phi}{\partial\rho}+\frac{1}{\rho^{2}}\left(\frac{\partial}{\partial\varphi}-\chi\frac{\partial}{\partial z}\right)^{2}\phi+\frac{\partial^{2}\phi}{\partial z^{2}}.
\label{3.4}
\end{eqnarray}
The solution to Eq. (\ref{3.4}) has the same form of Eq. (\ref{2.5}), therefore, we obtain the following radial equation:
\begin{eqnarray}
f''+\frac{1}{r}\,f'-\frac{\left(l-\chi\,k\right)^{2}}{r^{2}}\,f+\lambda^{2}\,f=0,
\label{3.5}
\end{eqnarray}
where $\lambda$ has been defined in Eq. (\ref{2.7}). Note that Eq. (\ref{3.5}) is also the Bessel differential equation \cite{abra}. By following the steps from Eq. (\ref{2.8}) to Eq. (\ref{2.10}), we have that $f\left(r\right)=A\,J_{\left|l\right|}\left(\lambda\,r\right)$ and the boundary condition (\ref{2.9}) is replaced with
\begin{eqnarray}
f\left(r\rightarrow \bar{\rho}_{0}=\frac{\sqrt{1-\omega^{2}\chi^{2}}}{\omega}\right)=0,
\label{3.6}
\end{eqnarray}
which also corresponds to the scalar field subject to a hard-wall confining potential as in the previous section. Let us also consider $\lambda\,\bar{\rho}_{0}\gg1$, then, we obtain
\begin{eqnarray}
\mathcal{E}_{n,\,l,\,k}\approx -l\omega\pm \sqrt{m^{2}+\frac{\omega^{2}\,\pi^{2}}{\left(1-\omega^{2}\chi^{2}\right)}\left[n+\frac{1}{2}\,\left|l-\chi\,k\right|+\frac{3}{4}\right]^{2}+k^{2}}.
\label{3.10}
\end{eqnarray}

Hence, the relativistic energy levels (\ref{3.10}) are obtained when a scalar field is subject to a hard-wall confining potential under the effects of rotation in the spacetime with a space-like dislocation. The contributions to the relativistic energy levels (\ref{3.10}) that stem from the topology of the defect are given by the effective angular momentum $l_{\mathrm{eff}}=l-\chi\,k$, and by the presence of the fixed radius $\bar{\rho}_{0}=\frac{\sqrt{1-\omega^{2}\chi^{2}}}{\omega}$. Due to the effects of the torsion in the spacetime, we have a shift in the angular momentum that yields the effective angular momentum $l_{\mathrm{eff}}=l-\chi\,k$. As discussed in Refs. \cite{valdir3,vb}, this shift in the angular momentum corresponds to an analogue effect of the Aharonov-Bohm effect \cite{ab,pesk}. Note that for $\chi=0$, the contribution of torsion in the spacetime vanishes, and thus, we recover the relativistic energy levels (\ref{2.12}) in the Minkowski spacetime in a uniformly rotating frame. Again, we can also observe a Sagnac-type effect \cite{sag,r4,sag2,sag5} by the presence of the coupling between the angular velocity $\omega$ and angular momentum quantum number $l$.

\section{Rotating effects in the spacetime with a spiral dislocation}

In Ref. \cite{val}, examples of topological defects in solids associated with torsion are given. We start this section by considering a generalization of a topological defect in gravitation. Let us consider the distortion of a circle into a spiral, that corresponds to a spiral dislocation. Then, the corresponding line element of this topological defect in the context of gravitation is \cite{bf,val}:
\begin{eqnarray}
ds^{2}=-dt^{2}+dr^{2}+2\beta\,dr\,d\varphi+\left(\beta^{2}+r^{2}\right)d\varphi^{2}+dz^{2},
\label{4.1}
\end{eqnarray}
where $\beta$ is a constant ($\beta>0$) associated with the distortion of the defect.

\subsection{Scalar field subject to a hard-wall confining potential }

Let us investigate the topological effects of the spacetime with a spiral dislocation on the scalar field when it is subject to a hard-wall confining potential. The Klein-Gordon equation (\ref{2.2}) in the spacetime described by the line element (\ref{4.1}) is given by
\begin{eqnarray}
m^{2}\,\phi=-\frac{\partial^{2}\phi}{\partial t^{2}}+\left[1+\frac{\beta^{2}}{r^{2}}\right]\frac{\partial^{2}\phi}{\partial r^{2}}+\left[\frac{1}{r}-\frac{\beta^{2}}{r^{3}}\right]\frac{\partial\phi}{\partial r}-\frac{2\beta}{r^{2}}\frac{\partial^{2}\phi}{\partial r\,\partial\varphi}+\frac{\beta}{r^{3}}\frac{\partial\phi}{\partial\varphi}+\frac{1}{r^{2}}\,\frac{\partial^{2}\phi}{\partial\varphi^{2}}+\frac{\partial^{2}\phi}{\partial z^{2}}.
\label{4.2}
\end{eqnarray}
The solution to Eq. (\ref{4.2}) has the same form of Eq. (\ref{2.5}), therefore, from Eq. (\ref{4.2}) we obtain the radial equation:
\begin{eqnarray}
\left[1+\frac{\beta^{2}}{r^{2}}\right]f''+\left[\frac{1}{r}-\frac{\beta^{2}}{r^{3}}-\frac{i\,2\beta\,l}{r^{2}}\right]f'-\frac{l^{2}}{r^{2}}\,f+\frac{i\,\beta\,l}{r^{3}}\,f+\theta^{2}\,f=0,
\label{4.3}
\end{eqnarray}
where $\theta^{2}=\mathcal{E}^{2}-m^{2}-k^{2}$. With the purpose of solving this radial equation, let us write
\begin{eqnarray}
f\left(r\right)=\exp\left(i\,l\,\tan^{-1}\left(\frac{r}{\beta}\right)\right)\times u\left(r\right),
\label{4.4}
\end{eqnarray}
then, we obtain the following equation for $u\left(r\right)$:
\begin{eqnarray}
\left[1+\frac{\beta^{2}}{r^{2}}\right]u''+\left[\frac{1}{r}-\frac{\beta^{2}}{r^{3}}\right]u'-\frac{l^{2}}{\left(\beta^{2}+r^{2}\right)}\,u+\theta^{2}\,u=0.
\label{4.5}
\end{eqnarray}
We proceed with a change of variables given by $x=\theta\,\sqrt{r^{2}+\beta^{2}}$, and thus, we obtain the following equation:
\begin{eqnarray}
u''+\frac{1}{x}\,u'-\frac{l^{2}}{x^{2}}\,u+u=0.
\label{4.6}
\end{eqnarray}
Hence, Eq. (\ref{4.6}) is the Bessel differential equation \cite{abra}. We also follow the steps from Eq. (\ref{2.8}) to Eq. (\ref{2.10}), then, we can write $u\left(x\right)=A\,J_{\left|l\right|}\left(x\right)$ and the boundary condition (\ref{2.9}) is replaced with
\begin{eqnarray}
u\left(x\rightarrow x_{0}=\theta\sqrt{r^{2}_{0}+\beta^{2}}\right)=0,
\label{4.7}
\end{eqnarray}
which also corresponds to the scalar field subject to a hard-wall confining potential ($r_{0}$ is a fixed radius). Let us also consider $x_{0}\gg1$, and then,
by using the relation (\ref{2.10}), we obtain
\begin{eqnarray}
\mathcal{E}_{n,\,l,\,k}\approx\pm \sqrt{m^{2}+k^{2}+\frac{\pi^{2}}{\left(r_{0}^{2}+\beta^{2}\right)}\left[n+\frac{\left|l\right|}{2}+\frac{3}{4}\right]^{2}}.
\label{4.8}
\end{eqnarray}

Therefore, we have obtained in Eq. (\ref{4.8}) the relativistic spectrum of energy for the scalar field subject to the hard-wall confining potential in the spacetime with a spiral dislocation. The effect of the topology of this spacetime can be viewed by the presence of the parameter $\beta$ that gives rise to an effective radius $\zeta_{0}=\sqrt{r_{0}^{2}+\beta^{2}}$. In contrast to Eqs. (\ref{2.11}) and (\ref{3.10}), we have that the angular momentum remains unchanged, i.e., there is no effect of the topology of the spacetime that yields an effective angular momentum. Hence, there is no analogue of the Aharonov-Bohm effect for bound states \cite{pesk,valdir2,fur4,ab} in this sense. Note that, by taking $\beta=0$ we obtain the spectrum of energy in the Minkowski spacetime.

\subsection{Rotating effects on the scalar field subject to a hard-wall confining potential }

Henceforth, let us consider a uniformly rotating frame as in the previous sections. For this purpose, we also perform the coordinate transformation $\varphi\rightarrow\varphi+\omega\,t$, and then, the line element (\ref{4.1}) becomes:
\begin{eqnarray}
ds^{2}&=&-\left(1-\omega^{2}\,\beta^{2}-\omega^{2}\,r^{2}\right)dt^{2}+2\beta\,\omega\,dr\,dt+2\omega\left(\beta^{2}+r^{2}\right)d\varphi\,dt\nonumber\\
[-2mm]\label{4.9}\\[-2mm]
&+&dr^{2}+2\beta\,dr\,d\varphi+\left(\beta^{2}+r^{2}\right)d\varphi^{2}+dz^{2}.\nonumber
\end{eqnarray}
In this case, we have that the radial coordinate is restricted by range:
\begin{eqnarray}
0\,\leq\,r\,<\frac{\sqrt{1-\beta^{2}\omega^{2}}}{\omega}.
\label{4.10}
\end{eqnarray}
Hence, the restriction on the radial coordinate is determined by the angular velocity and the parameter associated with the distortion of a circle into a spiral (torsion). This restriction on the radial coordinate differs from the cosmic string spacetime given in Eq. (\ref{2.3a}), but it is analogous to that of the spacetime with space-like dislocation given in Eq. (\ref{3.3}). Therefore, if $r\,\geq\,\frac{\sqrt{1-\beta^{2}\omega^{2}}}{\omega}$. we would have a particle is placed outside of the light-cone. Note that by taking $\beta=0$, we also recover the discussion made in Ref. \cite{landau3} in the Minkowski spacetime.

Further, the Klein-Gordon equation (\ref{2.2}) in the spacetime described by the line element (\ref{4.9}) is given by
\begin{eqnarray}
m^{2}\,\phi&=&-\frac{\partial^{2}\phi}{\partial t^{2}}+2\omega\,\frac{\partial^{2}\phi}{\partial\varphi\,\partial t}+\left[1+\frac{\beta^{2}}{r^{2}}\right]\frac{\partial^{2}\phi}{\partial r^{2}}+\left[\frac{1}{r}-\frac{\beta^{2}}{r^{3}}\right]\frac{\partial\phi}{\partial r}-\frac{2\beta}{r^{2}}\frac{\partial^{2}\phi}{\partial r\,\partial\varphi}\nonumber\\
[-2mm]\label{4.11}\\[-2mm]
&+&\frac{\beta}{r^{3}}\frac{\partial\phi}{\partial\varphi}+\left[\frac{1}{r^{2}}-\omega^{2}\right]\frac{\partial^{2}\phi}{\partial\varphi^{2}}+\frac{\partial^{2}\phi}{\partial z^{2}}.\nonumber
\end{eqnarray}
The solution to Eq. (\ref{4.11}) has the same form of Eq. (\ref{2.5}), therefore, the radial equation becomes
\begin{eqnarray}
\left[1+\frac{\beta^{2}}{r^{2}}\right]f''+\left[\frac{1}{r}-\frac{\beta^{2}}{r^{3}}-\frac{i\,2\beta\,l}{r^{2}}\right]f'-\frac{l^{2}}{r^{2}}\,f+\frac{i\,\beta\,l}{r^{3}}\,f+\lambda^{2}\,f=0,
\label{4.12}
\end{eqnarray}
where we have also defined $\lambda$ in Eq. (\ref{2.7}). By following the steps from Eq. (\ref{4.4}) to Eq. (\ref{4.6}), but with the change of variables given by $y=\lambda\,\sqrt{r^{2}+\beta^{2}}$, we obtain 
\begin{eqnarray}
u''+\frac{1}{y}\,u'-\frac{l^{2}}{y^{2}}\,u+u=0.
\label{4.13}
\end{eqnarray}
Hence, by following the steps from Eq. (\ref{2.8}) to Eq. (\ref{2.10}), then, we have that $u\left(y\right)=A\,J_{\left|l\right|}\left(y\right)$. With $\rho_{0}'=\frac{\sqrt{1-\beta^{2}\omega^{2}}}{\omega}$, therefore, the boundary condition (\ref{2.9}) becomes
\begin{eqnarray}
u\left(y\rightarrow y_{0}= \frac{\lambda}{\omega}\right)=0,
\label{4.14}
\end{eqnarray}
which also corresponds to the scalar field subject to a hard-wall confining potential as in the previous sections. Let us also consider $y_{0}\gg1$, then, we obtain
\begin{eqnarray}
\mathcal{E}_{n,\,l,\,k}\approx -l\omega\pm \sqrt{m^{2}+k^{2}+\omega^{2}\,\pi^{2}\left[n+\frac{\left|l\right|}{2}+\frac{3}{4}\right]^{2}}.
\label{4.14}
\end{eqnarray}

In Eq. (\ref{4.14}), we have the relativistic spectrum of energy for the scalar field subject to a hard-wall confining potential under the effects of rotation in the spacetime with a spiral dislocation. In contrast to Eqs. (\ref{2.11}), (\ref{3.10}) and (\ref{4.8}), there is no influence of the topology of the spacetime on the relativistic energy levels in the uniformly rotating frame. Despite having effects of the topology of the spacetime on the radial coordinate as we have seen in Eq. (\ref{4.10}), for $\rho_{0}'=\frac{\sqrt{1-\beta^{2}\omega^{2}}}{\omega}$, we have only effects of rotation on the relativistic energy levels. The Sagnac-type effect \cite{sag,r4,sag2,sag5} can be observed in Eq. (\ref{4.14}) due to the presence of the coupling between the angular velocity $\omega$ and angular momentum quantum number $l$. Besides, there is no effect analogous to the Aharonov-Bohm effect for bound states \cite{pesk,valdir2,fur4,ab}, since the angular momentum remains unchanged. Therefore, the spectrum of energy (\ref{4.14}) is analogous to that obtained in Eq. (\ref{2.12}) in the Minkowski spacetime.

\section{conclusions}

We have investigated rotating effects on the scalar field confined to a hard-wall confining potential in three different topological defect spacetimes: the cosmic string spacetime, the spacetime with a space-like dislocation and the spacetime with a spiral dislocation. We have started our discussion in all these cases through the restriction on the radial coordinate that arises from the uniformly rotating frame and the topology of the spacetime. Then, we have used this information about the restriction on the radial coordinate to impose the boundary condition that corresponds to a hard-wall confining potential. We have shown in all these cases that a discrete spectrum of energy can be achieved. In the case of the cosmic string spacetime, we have seen that the spectrum of energy depends on the fixed radius $\rho_{0}=\frac{1}{\alpha\omega}$ and the effective angular momentum $l_{\mathrm{eff}}=\frac{\left|l\right|}{\alpha}$. On the other hand, in the case of the spacetime with a space-like dislocation, we have seen a dependence of the spectrum of energy on the fixed radius $\bar{\rho}_{0}=\frac{\sqrt{1-\omega^{2}\chi^{2}}}{\omega}$ and the effective angular momentum $l_{\mathrm{eff}}=l-\chi\,k$. Hence, due to this presence of the effective angular momentum, we have an analogue of the Aharonov-Bohm effect for bound states \cite{pesk,valdir2,fur4,ab}. Furthermore, in these two cases we have also seen a Sagnac-type effect \cite{sag,r4,sag2,sag5}.

The spacetime with a spiral dislocation corresponds to a generalization of a topological defect (the distortion of a circle into a spiral \cite{val}) in gravitation. With this background, we have first analysed the confinement of the scalar field to a hard-wall confining potential without rotating effects. We have seen that a discrete spectrum of energy can also be obtained in this topological defect background, where the topology of the spacetime yields a contribution that gives rise to an effective radius $\zeta_{0}=\sqrt{r_{0}^{2}+\beta^{2}}$. On the other hand, no analogue effect of the Aharonov-Bohm effect for bound states exists. In the second case analysed, we have considered the uniformly rotating frame. We have also seen that there exists a restriction on the radial coordinate that arises from the uniformly rotating frame and the topology of the spacetime. By investigating the confinement of the scalar potential to a hard-wall confining potential, we have also obtained a discrete spectrum of energy. In this case, even though there exists the presence of the fixed radius $\rho_{0}'=\frac{\sqrt{1-\beta^{2}\omega^{2}}}{\omega}$ which determines the boundary condition of the hard-wall confining potential, we have seen no dependence of the relativistic energy levels on the parameter associated with the spiral dislocation spacetime. Besides, no analogue effect of the Aharonov-Bohm effect for bound states exists. Due to the effects of rotation, a contribution to the relativistic energy levels that gives rise to a Sagnac-type effect exists. Therefore, the relativistic energy levels obtained in the spacetime with a spiral dislocation in a uniformly rotating frame are analogous to the case of the Minkowski spacetime.

\acknowledgments

The authors would like to thank the Brazilian agencies CNPq and CAPES for financial support.

\end{document}